\documentclass{revtex4}
\usepackage{amsmath}
\usepackage{amsmath,amssymb,amsthm}
\usepackage{epsfig}
\usepackage{epstopdf}
\usepackage{graphicx}

\usepackage{verbatim}

\newcommand{\mapleplot}[4]{%
\begin{figure}[htb!]
\includegraphics[scale={#4}]{#1}
\caption{#2}
\label{#3}
\end{figure}}

\begin{document}

\title{Quasi-Topological Lifshitz Black Holes}
\author{W. G. Brenna$^{1,2}$\footnote{email address: wgb278@mail.usask.ca}, M. H. Dehghani$^{3,4}$\footnote{email address: mhd@shirazu.ac.ir}, and R. B. Mann$^{1}$\footnote{email address: rbmann@sciborg.uwaterloo.ca}}
\affiliation{$^1$Department of Physics \& Astronomy, University of Waterloo, 200 University Avenue West, Waterloo, Ontario, Canada, N2L 3G1\\$^2$Department of Physics and Engineering Physics, University of Saskatchewan, 116 Science Place, Saskatoon, SK, S7N 5E2\\$^3$Physics Department and Biruni Observatory, College of Sciences, Shiraz University, Shiraz 71454, Iran\\ $^4$ Research Institute for Astrophysics and Astronomy of Maragha
(RIAAM), Maragha, Iran}
\date{\today}

\begin{abstract}

We investigate the effects 
 of including a quasi-topological cubic curvature term to the Gauss-Bonnet action to five dimensional  Lifshitz gravity.
We find that a new set of Lifshitz black hole solutions exist that are analogous to those obtained in third-order Lovelock gravity
in higher dimensions.  No additional matter fields are required to obtain solutions with asymptotic Lifshitz behaviour, though
we also investigate solutions with matter. 
Furthermore, we examine black hole solutions and their thermodynamics in this situation and find that a
 negative quasi-topological term, just like a positive Gauss-Bonnet term, prevents instabilities in what are ordinarily unstable Einsteinian black holes.

\end{abstract}

\maketitle

\section{Introduction}

The concept of holography has proven to be enormously fruitful for
demonstrating interesting new connections between disparate areas of physics.
The basic idea is that gravitational dynamics in a given dimensionality can be
mapped onto some other (non-gravitational) field theory of a lower dimensionality.
Holography has been most thoroughly explored in the context of the AdS/CFT
correspondence conjecture, in which a large volume of calculational evidence
indicates that a (relativistic) conformal field theory (CFT)
can be mapped to gravitational dynamics in an asymptotically Anti de Sitter (AdS) spacetime
of one larger dimension \cite{AdSCFT}.

Over the past few years it has become clear that holographic concepts cover a much
broader conceptual territory. For example holographic renormalization has been shown
to be a useful tool for understanding conserved quantities and gravitational thermodynamics
in both asymptotically de Sitter \cite{dSCFT} and asymptotically flat spacetimes \cite{MannMarolf}.
Much more recently holography has been extended to describe a duality
between a broad range of strongly coupled field theories and gravity
in the context of QCD quark-gluon plasmas \cite{Kov}, atomic physics, and condensed matter physics \cite{Hart,Faul,Mc}.

Gravity-gauge duality is evidently a robust concept, and its full implications for physics
(for example in elucidating the strong coupling behaviour of the non-gravitational theories
noted above) remain  to be understood. One line of investigation has been concerned with
Lifshitz field theories, which have an anisotropic scaling of the form
\begin{equation}
t \rightarrow \lambda^z t, \hspace{5mm} r \rightarrow \lambda^{-1} r, \hspace{5mm} x \rightarrow \lambda x
\end{equation}
exhibited by fixed points governing the behaviour of various condensed matter systems.
Such scaling offers some promise for further extending holographic duality between condensed matter physics and gravity. While for $z=1$ this scaling symmetry is the familiar conformal
symmetry, for $z=3$, theories with this type of scaling are
power-counting renormalizable, possibly providing a UV completion to the effective gravitational field theory \cite{Horava}. For Lifshitz gravity, the natural asymptotic spacetime metric is
\begin{equation}
ds^{2}= - \frac{r^{2z}}{L^{2z}}dt^{2}+ L^{-2} \left( \frac{dr^{2}}{r^{2}}+r^{2}d\Omega%
^{2}\right)  \label{Lifmet}
\end{equation}
noted earlier in a braneworld context \cite{Koroteev}.

A $D$-dimensional anisotropic scale invariant background using an action
that couples gravity to a massive gauge field (or alternatively to 2-form
and dualized $(D-1)$-form field strengths with a Chern-Simons coupling) can be constructed
that has solutions with the asymptotic behaviour (\ref{Lifmet}) \cite
{Kach}. An early example \cite{Tay} for an extended class of vacuum solutions
for a sort of higher-dimensional dilaton gravity with general $z$ was soon followed
by the discovery of black hole solutions, both exact (for $z=2$) \cite{Mann} and
numerical (for more general values of $z$) \cite{Mann,Peet,Bal,Dan}.

Since in general one expects quantum-gravitational effects to induce corrections
to the Einstein action, it is natural to consider modifying the gravitational part of
the action with higher-derivative terms due to additional powers of the curvature.
Such terms must be considered on the gravity side of the duality conjecture
 in order to study CFTs with different values for their central charges. Here,
 Lovelock gravity theories play a special role in that the number of metric
derivatives in any field equation is never larger than 2. Furthermore, third-order Lovelock gravity is supersymmetric,
and therefore one can define superconformal field theories via the AdS/CFT
correspondence \cite{Boer,Cam}.

The addition to the action of a term  cubic in curvature is not new, but asymptotic Lifshitz solutions
in Lovelock gravity coupled to a massive Abelian gauge field
were only recently discovered \cite{Deh2}. For a suitable choice of coupling constant one can dispense
with this massive gauge field, since the additional Lovelock terms can play the role of the desired matter. Some new exact black hole solutions were obtained as well as a broad class of numerical solutions, and asymptotic Lifshitz solutions with curvature-squared terms in the action have also been investigated  \cite{Ay1,Ay2,Cai,Pang}. Somewhat remarkably, the relationship between the energy density, temperature, and entropy density is unchanged from Einsteinian gravity \cite{Deh3} even though
the sub-leading large-$r$ behaviour of Lovelock-Lifshitz black branes differs substantively from their Einsteinian Lifshitz counterparts \cite{Mann,Dan}. The relationship between entropy and temperature is also the same as the Einsteinian case, apart from a constant of integration that depends on the Lovelock coefficients.

The hallmark feature of Lovelock theories is that no field equation has more than two derivatives of any metric coefficient. However, a new cubic curvature term was recently introduced that can perhaps be regarded as a generalization of Lovelock gravity in five dimensions \cite{Oliva2,Myers}.
The generality arises from a spherical symmetry requirement: the field equations will generally reduce to second-order system of differential equations when the metric is spherically symmetric.

This particular class of correction terms has been coined quasi-topological gravity, since in some ways they behave like topological invariants in 6 dimensions, yet for nonspherical geometries, they contribute nontrivially to the action. Furthermore, there are no Lagrangians that are cubic in curvature in four dimensions for spherical symmetry that lead  to second order differential equations.
Quasi-topological gravity has been previously studied in the case of planar AdS black holes.
In this paper we investigate the implications of this new term for asymptotically Lifshitz spacetimes.

Specifically, we examine the effects of higher-curvature modifications to Einsteinian gravity to asymptotically Lifshitz metrics, both with and without massive background Abelian gauge fields.
We find that quasi-topological Lifshitz gravity replicates the field equations from third-order Lovelock-Lifshitz gravity \cite{Deh2}, provided the quasi-topological parameter $\mu$ is appropriately renormalized.
We find that indeed, asymptotic Lifshitz black holes exist in both cases.  We obtain both exact solutions and numerical ones,
the latter obtained via the shooting method.
We close with a short discussion of the relevant thermodynamics and conserved quantities of our black hole solutions.

\section{Quasi-Topological Gravity \label{Fieldequations}}

The quasi-topological additions consist of 3rd-order curvature corrections to Gauss-Bonnet gravity that maintain second-order field equations with respect to the metric under conditions of spherical symmetry.
We use the action
\begin{equation}
I = \int d^{D} x \sqrt{-g} \left( - 2 \Lambda + \mathcal{L}_1 + \frac{\lambda L^2}{(D-3)(D-4)}\mathcal{L}_2 - \frac{8 \mu L^4}{(D-3)(D-6)}\mathcal{L}_3 - \frac{1}{4}F_{\mu \nu}F^{\mu \nu} -\frac{1}{2}m^2A_{\mu}A^{\mu}   \right)
\label{action}
\end{equation}
where $D$ is the number of dimensions (larger than 4 and different from 6), $F_{\mu \nu} = \partial_{[\mu}A_{\nu]}$, $\mu$ and $\lambda$ are the correction terms' coefficients, $\mathcal{L}_1 = R$ is the Ricci scalar, $\mathcal{L}_2 = R_{\mu \nu \gamma \delta} R^{\mu \nu \gamma \delta} - 4 R_{\mu \nu} R^{\mu \nu} + R^2$ is the Gauss-Bonnet Lagrangian, and $\mathcal{L}_3$ is the quasi-topological gravity correction.
This quasi-topological gravity correction has the form
\begin{align}
\mathcal{L}_3 &= \frac{2D-3}{3D^2-15D+16}{{{R_\mu}^\nu}_\lambda}^\rho {{{R_\nu}^\tau}_\rho}^\sigma {{{R_\tau}^\mu}_\sigma}^\lambda + \frac{3}{(D - 4)(3D^2-15D+16)} \left( \frac{3D - 8}{8} R_{\mu \lambda \nu \rho} R^{\mu \lambda \nu \rho} R \right. \nonumber \\
              & \quad \left. - (D-2) R_{\mu \lambda \nu \rho} {R^{\mu \lambda \nu}}_\tau R^{\rho \tau} + D R_{\mu \lambda \nu \rho} R^{\mu \nu}R^{\lambda \rho} + 2(D-2) {R_\mu}^\lambda {R_\lambda}^\nu {R_\nu}^\mu - \frac{3D-4}{2} {R_\mu}^\lambda {R_\lambda}^\mu R + \frac{D}{8} R^3 \right)
\label{quasitop}
\end{align}
This term is only effective in dimensions greater than 4, and it becomes trivial in 6 dimensions \cite{Myers}.

Rather than write down the full tensorial expression for the field equations, as we are interested only in spherically symmetric solutions we will insert the Lifshitz metric
\begin{equation}\label{metlif}
ds^2 = -\frac{r^{2 z}}{L^{2 z}} f(r) dt^2 + \frac{L^2 dr^2}{r^2 g(r)} +   r^2  d\Omega^2
\end{equation}
into the action and then functionally vary it, obtaining (after eliminating redundancies) three equations of motion for
the two metric functions and the gauge field.
Boundary conditions require that $f(r)$ and $g(r)$ asymptotically reach unity.
The term $d\Omega^2$ is the metric for a constant curvature hypersurface
\begin{equation}
d\Omega^2 =d{\theta_1}^2 + k^{-1}\sin^2 {\left(\sqrt{k} \theta_1\right)} \left( d{\theta_2}^2 + \displaystyle\sum\limits_{i=3}^{D-2%
	} \displaystyle\prod\limits_{j=2}^{i-1} \sin^2{\theta_j } d{\theta_i}^2 \right)
\end{equation}
where parameter $k$ is either $-1$, $0$, or  $1$, providing hyperbolic, flat, and spherical geometries, respectively. For $k=0$ a coordinate transformation will reduce this to the form $\sum_{k}^{D-2}d{\theta_k}^2$.
Symmetry requirements imply that the gauge field ansatz is
\begin{equation}\label{gfield}
A_t = q \frac{r^z}{L^z} h(r) .
\end{equation}
with all other components vanishing.

Now that the formalism has been specified, we restrict our considerations to five dimensions
and so (unless otherwise stated) the following results are only valid $D=5$.
Rather than carry out
a full variational principle, we insert
the ansatz (\ref{metlif}) and (\ref{gfield}) into the action, obtaining the effective action
\begin{equation}
\label{evalaction}
I = \int d^{4} x \int dr \frac{ r^{z-1} }{k L^{z+1}} \sqrt{\frac{f}{g}} \left( \left\{ 3 r^4 \left( \frac{- \Lambda}{6}L^2 %
	- \kappa + \lambda \kappa^2 + \mu \kappa^3 \right) \right\}^{\prime}+ \frac{q^2 r^3}{2 f} \left( g \left( r h^{\prime} + z h \right)^2 + m^2 L^2 h^2	\right) \right)
\end{equation}
for the spherically symmetric case, where $ \kappa = \left( g - \frac{L^2}{r^2} k\right) $.

Functionally varying (\ref{evalaction}) with respect $g(r)$, $f(r)$, and $h(r)$ respectively yields upon simplification
\begin{align}
& \Lambda L^2 r^6 + \left( 3z+3 \right) r^6 g - 6z \lambda r^6 g^2 + 6z \lambda r^4 L^2 k g - 3 r^4 L^2 k - \left( 9z-3 \right) \mu r^6 g^3 + \left( 18z-9 \right) \nonumber \\
& \mu r^4 L^2 k g^2 - \left(9z-9 \right) \mu L^4 k^2 r^2 g - 3 \mu L^6 k^3 + g (\ln{f})^{'} \left( \frac{3}{2} r^7 - 3 \right. \lambda r^7 g + 3 \lambda r^5 L^2 k - \frac{9}{2} \mu r^7 g^2 \nonumber \\
& + \left. 9 \mu r^5 g L^2 k - \frac{9}{2} \mu r^3 L^4 k^2 \right) = \frac{q^2 r^6}{4 f}\left[ g \left(r h^{'} + z h \right)^2 - m^2 L^2 h^2 \right] \label{fieldequations_initial}\\
& \left( 3 r^4 \left[ - \frac{\Lambda}{6} L^2%
- \kappa + \lambda \kappa^2 + \mu \kappa^3 \right] \right)^{\prime}%
= \frac{q^2 r^3}{2 f}\left[ g \left(r h^{'} + z h \right)^2 + m^2 L^2 h^2 \right]
\label{fieldequations_initial2}\\
&2r^{2}h^{\prime \prime }- r\left[ (\ln f)^{\prime }-(\ln%
g)^{\prime }\right]( rh^{\prime }+zh) +2(z+4) rh^{\prime }+6zh=2m^{2}L^{2}\frac{h}{g}
\label{fieldequations_final}
\end{align}
where a prime ($^\prime$) represents differentiation with respect to the radial coordinate $r$.

Before trying to find solutions to the above equations, we present a first integral
for the above equations of motion. It is a matter of calculation to show that this conserved quantity
can be written as
\begin{equation}
\mathcal{C}_0 = \left[ \left( 1 - 2 \lambda g - 3 \mu g^2 \right) \left( r f^{\prime} + 2 \left( z - 1 \right) f \right) - q^2 \left( z h + r h^{\prime} \right) h%
\right] \frac{r^{z+D-2}}{L^{z+1}} \left( \frac{f}{g} \right)^{1/2} .
\end{equation}
with details of this result given in appendix \ref{Conserved}.
For $z=1$, $f(r) = g(r)$ and the constant reduces (in the matter-free case) to
\begin{equation*}
\mathcal{C}_0 = \frac{r^{D}}{L^2} \left( f - \lambda f^2 - \mu f^3 \right)^{\prime},
\end{equation*}
which is proportional to the mass of black hole.

\section{Black Holes}

\subsection{Matter-free Solutions}

Setting $h(r)=0$, we first consider the existence of solutions of the form
\begin{equation}
ds^{2}=-\frac{r^{2z}}{L^{2z}}dt^{2}+\frac{L^2 dr^{2}}{r^{2}}+r^{2}\sum%
\limits_{i=1}^{3}d\theta _{i}^{2},  \label{met2}
\end{equation}
where $k=0$.  This is a Lifshitz analogue of flat space for asymptotically flat solutions, whose properties have been discussed
elsewhere \cite{Keith}. We shall refer to such solutions
as ``Lifshitz solutions''.

For the metric (\ref{met2}) the field  equations (\ref{fieldequations_initial}, \ref{fieldequations_initial2})
imply
\begin{equation}
\Lambda = - \frac{2}{L^2} \left( 2 - \lambda \right), \hspace{5mm}%
\mu = \frac{1}{3} \left( 1 - 2 \lambda \right) , \label{Con1}
\end{equation}
independent of our choice of z. These constraints reduce to those of five dimensional Gauss-Bonnet gravity \cite{Deh2}
\begin{equation}
\Lambda = -\frac{3}{L^2} \text{ \ \ and \ \ } \lambda = \frac{1}{2}
\end{equation}
when $\mu = 0$.
Note that the same constraints are necessary to ensure the existence of asymptotic Lifshitz solutions if $k \ne 0$.

With the above constraints, the exact Lifshitz solution (that is, $f(r) = g(r) =1$) is a solution to the field equations for any value of $z$.
Eq. (\ref{fieldequations_initial2}) with the condition (\ref{Con1}) reduces to
\begin{equation}\label{kapeq}
2-\lambda-3\kappa+3 \lambda \kappa^2 + (1-2 \lambda)\kappa^3=\frac{C}{r^4}
\end{equation}
where $C$ is a constant of integration. For $C=0$, $\kappa=1$ or
\begin{equation}
g(r) = 1 + \frac{k L^2}{r^2}
\end{equation}
yielding the only solution of eq. (\ref{kapeq}) that has the desired asymptotic behaviour.
The function $f(r)$ is not restricted by Eq. (\ref{fieldequations_initial}).
This degeneracy of the field equations has been noted previously in 5 dimensional Einstein-Gauss-Bonnet gravity with a cosmological constant \cite{Oliva} and third order Lovelock gravity \cite{Deh2}.
In the Gauss-Bonnet case, it was shown that there exists a degenerate set of solutions where $f(r)$ is left unspecified, while for certain values of the Gauss-Bonnet parameter, $f(r) = g(r)$.
In our case, this degeneracy is lifted when matter is present, and we obtain a family of solutions that become unique for a specific field strength, as we shall see.

Choosing $f(r) = g(r)$ (as in Lovelock gravity \cite{Deh2}) yields for
$k = -1$ an event horizon, and consequently the metric
\begin{equation}\label{metbhexact}
ds^2 = -\frac{r^{2 z}}{L^{2 z}} \left(1 - \frac{L^2}{r^2}\right) dt^2 + \frac{L^2 dr^2}{r^2 (1 -\frac{L^2}{r^2})} + r^2 d\Omega_{-1}^2
\end{equation}
which is an exact black hole solution.

For $C\neq 0$, one can find $\kappa$ and therefore $g(r)$, but upon inserting this solution in eq. (\ref{fieldequations_initial}), we find that the solution for $f(r)$ does not exhibit the desired asymptotic behaviour.
We find, with one exception, no other exact solutions to the field equations for these symmetries and asymptotic behaviour.

The exception is  $z=1$ (AdS), for which an exact solution can be found. The requirements that $f(r) = g(r)$ and $h(r)=0$ produce exact solutions dependent on $\lambda$ if $\mu=0$ \cite{Deh4}.
Setting $\mu\neq 0$, we first seek solutions for  $z=1$ without any background gauge field.
Restricting $f(r) = g(r)$ and setting $h(r) = 0$,
the field equation (\ref{fieldequations_final}) disappears, while the equations (\ref{fieldequations_initial}) and (\ref{fieldequations_initial2}) are not independent and can be analytically solved.
The result is
\begin{equation}\label{z1sol}
f(r) = g(r) = \frac{k L^2}{r^2} - \frac{\lambda}{3 \mu} +  \frac{1}{12 \mu r^2} \left[ \left( %
\sqrt{ \Gamma + J^2(r) } + J(r) \right)^{\frac{1}{3}} - \left( \sqrt{ \Gamma + J^2(r)} - J(r) \right)^{\frac{1}{3}} \right]
\end{equation}
where we define
\begin{equation}
\Gamma  = - \left( 16 r^4 \left( 3 \mu + \lambda^2 \right) \right)^3 \qquad
J(r)  = 16 r^6 \left(4 \lambda^3 + 18 \mu \lambda  -9 \mu^2 \Lambda L^2  - 18 \frac{M \mu^2}{r^4}   \right)
\end{equation}and $M$ is a constant of integration.
This solution matches the form of one obtained in 3rd order Lovelock gravity   for $D>6$ \cite{Deh4}, as our field equations are of the same form.
With this exact solution, we are able to compare results with the numerical algorithm.

\subsection{Matter solutions}

In the presence of a massive gauge field ($h(r)\neq 0$) the Lifshitz solution (\ref{met2})  is also supported by quasi-topological gravity  provided
\begin{align}
\label{lifshitzlovelockexact}
q^2 = \frac{2 \left( z - 1 \right) \left( 1 - 2\lambda - 3\mu \right)}{z} & &
m^2 = \frac{3 z}{L^2}   \nonumber \\
\Lambda = -\frac{1}{2L^2} \left[ ( 1 - 2\lambda - 3\mu ) ( 2z + z^2 ) + 9 - %
6\lambda - 3\mu \right] & &
\lambda < \frac{1}{2} ( 1 - 3\mu )
\end{align}
where the last constraint arises because we require $q^2 > 0$.  This in turn implies
$$
\frac{3\mu (z+1)^2-(z^2+2z +9)}{2L^2}  \leq \Lambda  \leq   -\frac{3}{2L^2} (1+\mu)
$$
provided $\lambda >0$, as is normative for Gauss-Bonnet gravity in the context of heterotic string theory \cite{Boul}.

We look for black hole solutions using both near-horizon and asymptotic series expansions of the metric and gauge functions.
The near-horizon series solutions are then used to obtain initial conditions for the numerical solution of the field equations.
The restrictions on $f(r)$ and $g(r)$ now become more rigid: these functions must not only approach unity as $r \to \infty$ (to satisfy the asymptotically Lifshitz boundary conditions) but they must also tend towards zero as $r \to r_0$ in order to ensure an event horizon exists.
First, we will show that series representations exist near and far from the horizon, and then we present a set of solutions obtained by numerically solving the differential equations (\ref{fieldequations_initial}-\ref{fieldequations_final}).

\subsubsection{Series Solutions}

We begin by searching for well-behaved black hole solutions in a near-horizon regime.
Our ansatz requires that the metric functions go to zero linearly near the horizon $r=r_0$:
\begin{align}
f(r)
&=f_{1}\left\{(r-r_{0})+f_{2}(r-r_{0})^{2}+f_{3}(r-r_{0})^{3}+...\right\}, \nonumber \\
g(r)
&=g_{1}(r-r_{0})+g_{2}(r-r_{0})^{2}+g_{3}(r-r_{0})^{3}+..., \label{near-hor}\\
h(r)
&=f_{1}^{1/2}\left\{h_{0}+h_{1}(r-r_{0})+h_{2}(r-r_{0})^{2}+h_{3}(r-r_{0})^{3}+...\right\}, \nonumber
\end{align}
and we find that the substitution of this ansatz into our equations of motion results in $h_0 = 0$ and a restriction on $g_1$:
\begin{align}\label{g1-nearhor}
g_1 = & \frac{z}{{r_0}^3} \left\{ 3 \mu \left( {r_0}^6 (z-1)^2 + 4z{r_0}^6 - 2L^6 k^3 \right) \right. \nonumber \\
& \quad \left. + 2 \lambda {r_0}^6 \left( (z+1)^2 + 2 \right) - {r_0}^4 \left( {r_0}^2 (z+1)^2 + 8 {r_0}^2 + 6L^2 k \right) \right\} \nonumber \\
& \quad \left[ 3 \mu \left(  (1-z) {h_1}^2 {r_0}^5 - 3 L^4 k^2 z \right) + 2 \lambda \left( (1-z) {r_0}^5 {h_1}^2  \right. \right. \nonumber \\
& \quad \left. \left. + 3 z L^2 k {r_0}^2 \right) + \left( (z-1) {h_1}^2 {r_0}^5 + 3 z {r_0}^4 \right) \right]^{-1} .
\end{align}

We are left with two free parameters, $h_1$ and $f_1$, and values for these are selected to ensure proper asymptotic behaviour for large $r$.
All of the other terms in the expansion are solvable in terms of these two parameters.

Solutions at large $r$ can be obtained by linearizing the system, using the ansatz
\begin{align}
f(r) &=1+\varepsilon f_{e}(r), \nonumber \\
g(r) &=1+\varepsilon g_{e}(r), \nonumber \\
h(r) &=1+\varepsilon h_{e}(r),
\label{larger_ansatz}
\end{align}
yielding rather lengthy expressions for the leading terms. We have relegated these to Appendix \ref{larger}.

\subsubsection{Numerical Solutions}
\label{numsols}
We will find it easier to obtain numerical solutions by writing 
\begin{equation}\label{hder}
\frac{d h}{d r} \equiv j(r) ,
\end{equation}
in which case the set of differential equations (\ref{fieldequations_initial}-\ref{fieldequations_final}) can be written as
\begin{align}
\frac{dj}{dr} &= \frac{ z h \left( 3 - 2 g \right) - z^2 h g - r j g \left( 2 z + 3 \right)}{g r^2} - %
L_0 \frac{ r^2 h^2 \left( z h + r j \right) \left( z - 1 \right)}{f g H} \notag \\
\frac{df}{dr} &= \frac{1}{ 3 zr^{3}gH}\Big\{3[4+6(z-1)]z\mu%
r^{6}fg^{3} \notag\\
& +3zr^{4}\left[ -3k\mu L^2 (4z-2)+(4z)%
\lambda r^{2}\right] fg^{2}  \notag \\
&-3zr^{2}\left[ 3(2z-2)k^{2}\mu+4zk
\lambda L^2 r^{2}+(2z+2)r^{4}\right] fg  \notag \\
&-z\mu\left\{[3(z-1)^{2}+12z]r^{6}-6k^3L^{6}\right\}f  \notag \\
&-z\lambda r^{2}\left\{
[2(z-1)^{2}+8z+4]r^{4}\right\} f  \notag \\
&+zr^{4}\left\{ [(z-1)^{2}+4z+8]r^{2}+6kL^{2}\right\} f
\notag \\
&+(z-1){L_0}r^{6}[(zh+rj)^{2}g-3zh^{2}]\Big\} \notag \\
\frac{dg}{dr} &= \frac{1}{3zr^{3}fH}\Big\{ 12z \mu%
r^{6}fg^{3}+3zr^{4}\left[ -6L^2k\mu+4\lambda%
r^{2}\right] fg^{2}  \notag \\
&-3zr^{2}\left[ 4k\lambda%
L^2r^{2}+4r^{4}\right] fg  \notag \\
&-z\mu\left\{[3(z-1)^{2}+12z]r^{6}-6k^3L^{6}\right\}f  \notag \\
&-z\lambda r^{2}\left\{
[2(z-1)^{2}+8z+4]r^{4}\right\} f  \notag \\
&+zr^{4}\left\{ [(z-1)^{2}+4z+8]r^{2}+6kL^{2}\right\} f
\notag \\
&+(z-1){L_0}r^{6}[(zh+rj)^{2}g+3zh^{2}]\Big\}, \label{dgr}
\end{align}
where for simplicity we define $L_0 = -1 + 2\lambda + 3\mu$ and $H = r^4 + 2 \lambda r^2 \left( k L^2 - r^2 g \right) - 3 \mu \left( k L^2 - r^2 g \right)^2$.

Equation  (\ref{hder}-\ref{dgr}) form a system of four coupled first order ordinary differential equations.
With these ODEs, initial conditions are chosen from the series solution (evaluated just beyond the horizon), and then the shooting method (explained in \cite{Mann,Dan}) is used to obtain solutions.

We consider values of $\mu$ and $\lambda$   that guarantee positivity of the energy flux in the dual conformal field theory \cite{Myers2} when $z=1$. For $z\neq 1$ the dual theory is not well understood and the analogous allowed ranges of $\mu$ and $\lambda$ are not known. Furthermore microscopic constraints such as  positivity of energy and causality 
are not necessarily responsible for setting the lower bound on the ratio of shear viscosity to entropy density in the plasma, since  hydrodynamic transport is determined by the infrared properties of the system, which do not necessarily enter into the microcausality analysis of the theory \cite{Sera}.  However for the most part we shall employ the same values of $\mu$ and $\lambda$  as for the $z=1$ case, noting departures from these values for illustrative purposes as appropriate.

The specific case  $z=1$ eliminates the charge $q$, and the solution is given by equation (\ref{z1sol}) with
$f(r)=g(r)$. Solving the system (\ref{fieldequations_initial}-\ref{fieldequations_final})
yields numerical solutions.

We can check the validity of our numerical approach by comparing this to the exact solution in equation (\ref{z1sol}).  For example, for $r_0 = 0.9$, and $k=0$ we see from Figure \ref{ancheck} that the two curves (numerical and analytic) are coincident.
To be certain,  we tested equality of the two approaches for $\mu = -0.001$, $\lambda = 0.04$, $k=0$, $r_0 = 1.5$.
Evaluating between $r=1.51$ and $r=15$ at intervals of $0.01$, we find that the two solutions differ by no more than $10^{-7}$.

\mapleplot{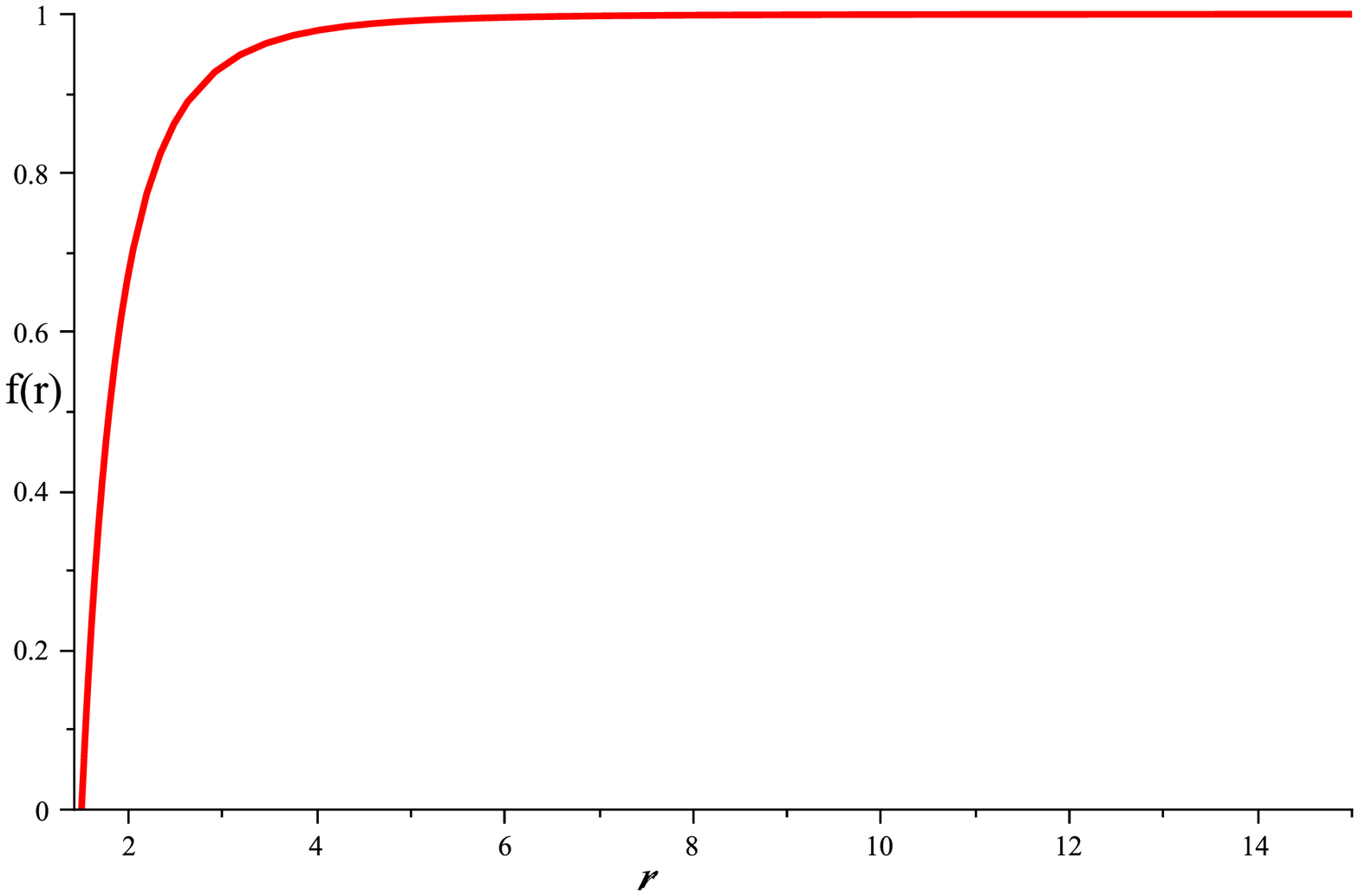}{Comparison of Analytic versus Numerical Solution for $f(r)$, where $r_0=0.9$, $\lambda=.04$, $\mu=-.001$, and $k=0$. The two curves are identical to one part in $10^{-7}$. }{ancheck}{0.4}

For $z \neq 1$, we numerically obtain solutions for large, medium, and small values of $r_0$ over a broad range of initial values of the field strength ($h_1$).
The quantity $f_1$ is then fixed by asymptotic conditions.
For a given value of $h_1$, we find that large black holes are asymptotic to functions that monotonically tend to unity, whereas the metric functions for small black holes exhibit a spike in magnitude before settling down.
However, due to the extra degree of freedom in the gauge field strength,  we can obtain a family of solutions (and control the spike) by varying  $h_1$, subsequently adjusting $f_1$ to satisfy the asymptotic conditions.
In Figure \ref{z2comp}, we see the result of varying the initial value of $h_1$ from $2.6$ (dashed solution) to $2.8$ (solid solution). For these the initial values of $f_1$ remained constant at $2.0$.
Note that the initial spike present for $h_1=2.6$ vanishes for $h_1=2.8 $.

In Figure \ref{z2plot1}, we plot the metric and gauge functions for a large black hole.
All three functions monotonically increase from zero at the horizon to unity for large r.
Figure \ref{z2plot3} shows a medium black hole ($r_0=2.4$), where the dashed line is Einsteinian gravity and the dotted line is quasi-topological gravity.
Here, due to the more favourable scale, we see that the solution for $h(r)$ is noticeably different.
The scale is still too large to see any effect on the $g(r)$ solution, however.

We can see from Figure \ref{z2plot2} that for small black holes, $f(r)$ spikes sharply.
The plot shows a comparison between Einsteinian gravity (dashed), Gauss-Bonnet gravity (solid), and quasi-topological gravity (dotted) for $k=-1$.
Small black holes for $k=0$ and $k=1$ exhibit similar behaviour.

The plot in Figure \ref{largemu} better shows the effect of larger values of $\mu$ and $\lambda$, elucidating how the quasi-topological term really affects solutions.

\mapleplot{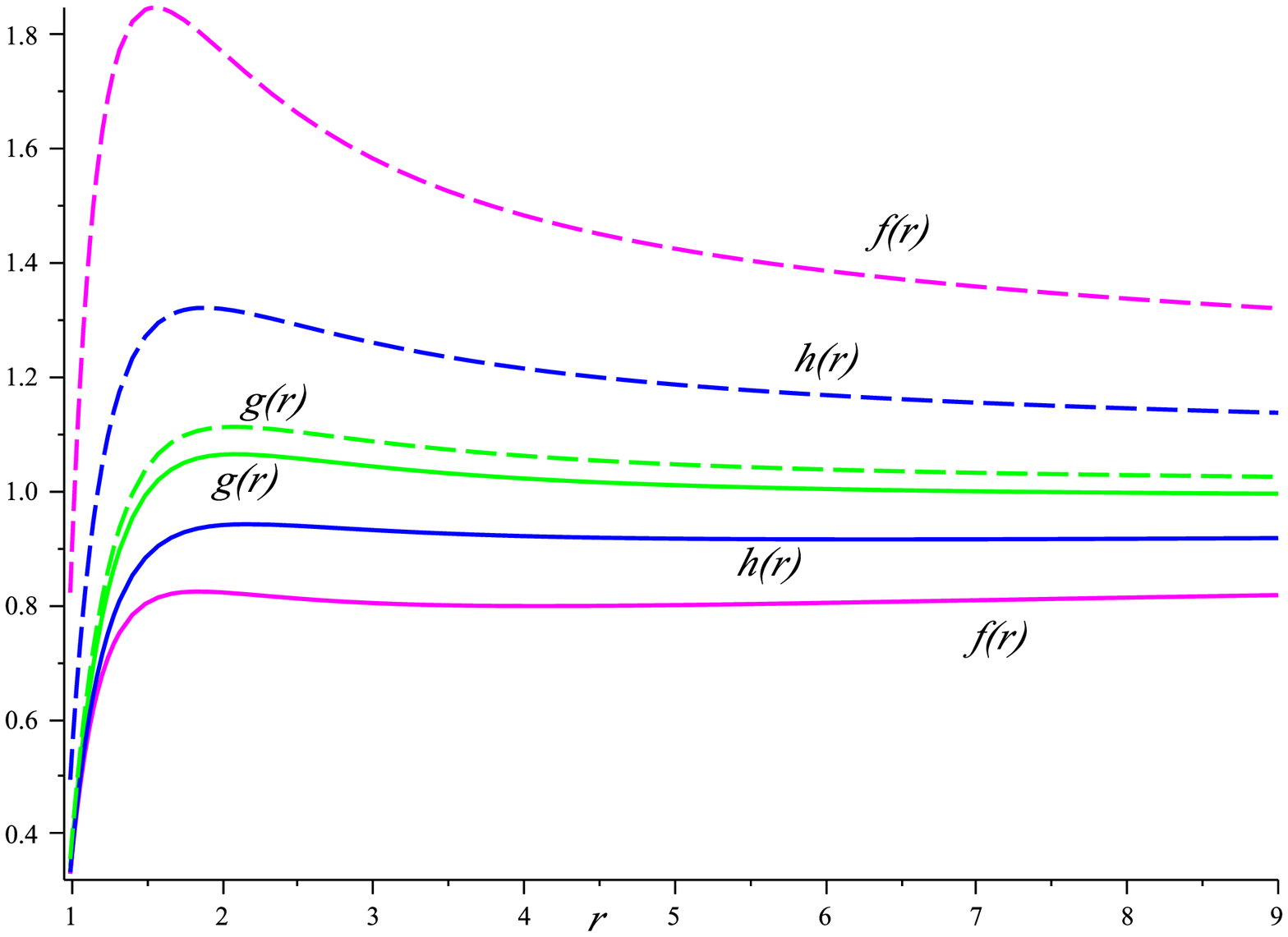}{Comparison of two sets of  $z=2$ solutions for $h_1=2.6$ (dashed) and $h_1=2.8$ (solid) for $\lambda=.1$ and $\mu=.001$, where $f(r), g(r), h(r)$ are plotted versus $r$ respectively in magenta, green and blue.}{z2comp}{0.4}
\mapleplot{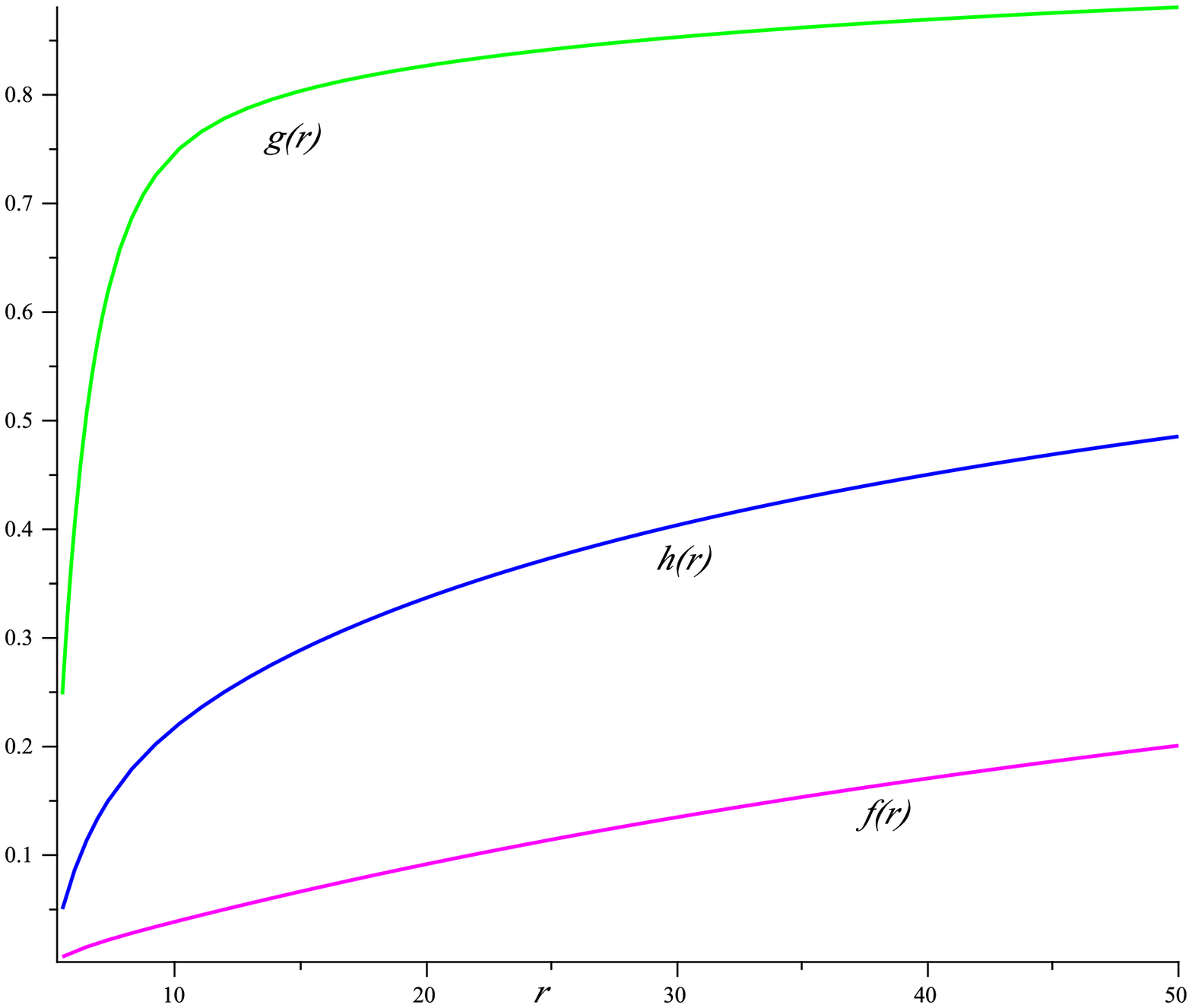}{Large black hole ($k=1$), where $\lambda=.04$ and $\mu=-.001$, with $f(r), g(r), h(r)$ versus $r$ respectively in magenta, green and blue for $z=2$. }{z2plot1}{0.4}
\mapleplot{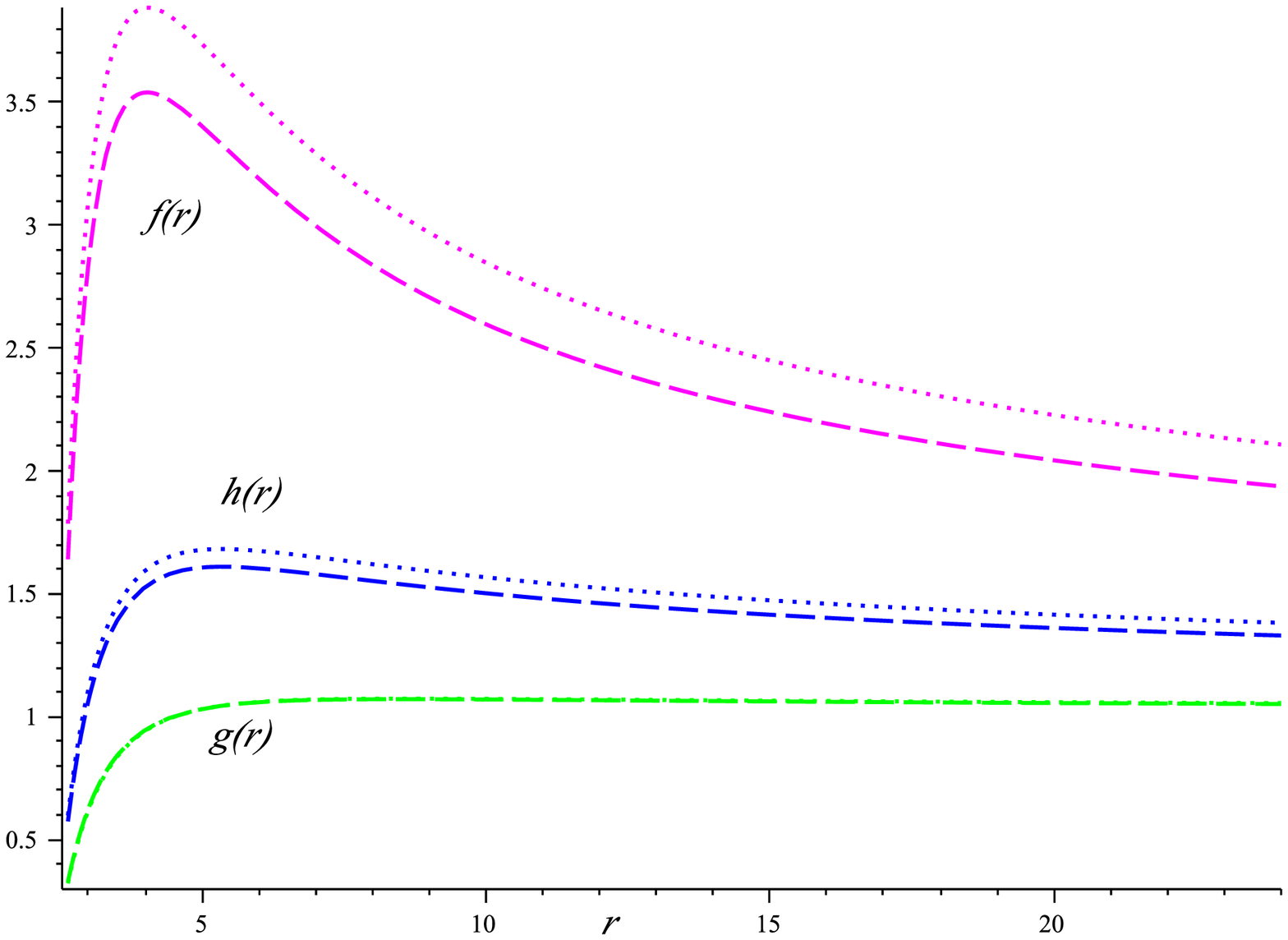}{Medium black hole of radius $r_0 = 2.4$ ($k=0$); here $\lambda=.04$ and $\mu=-.001$, and $f(r), g(r), h(r)$ versus $r$ for $z=2$ respectively in magenta, green and blue. The dotted line is Einsteinian gravity and the solid line is quasi-topological gravity.}{z2plot3}{0.4}
\mapleplot{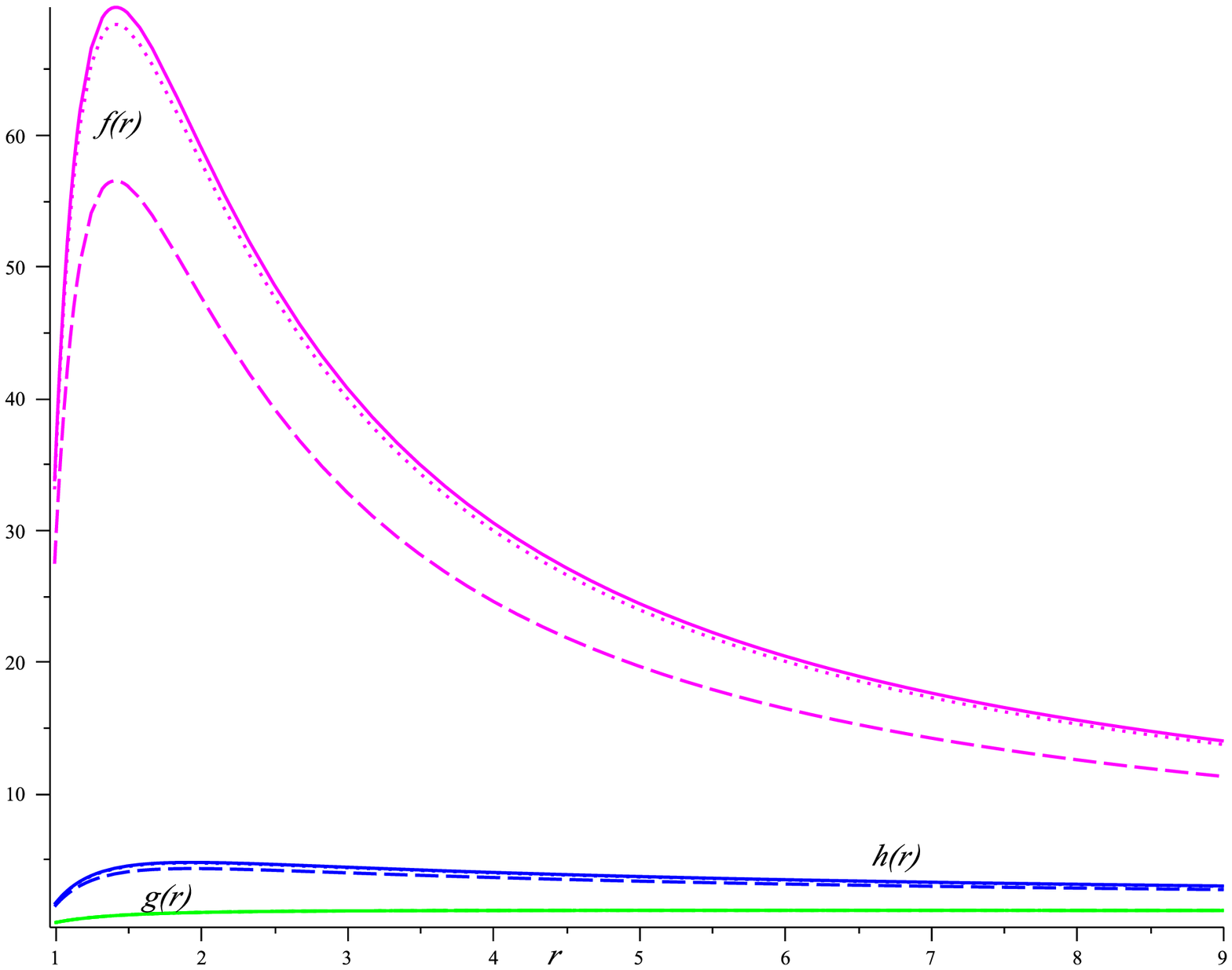}{Small black hole $r_0=0.9$ ($k=-1$), where $\lambda=.04$ and $\mu=-.001$, with $f(r), g(r), h(r)$ versus $r$ for $z=2$ respectively in magenta, green and blue for Einsteinian gravity (dashed), Gauss-Bonnet gravity (solid), and quasi-topological gravity (dotted). }{z2plot2}{0.4}
\mapleplot{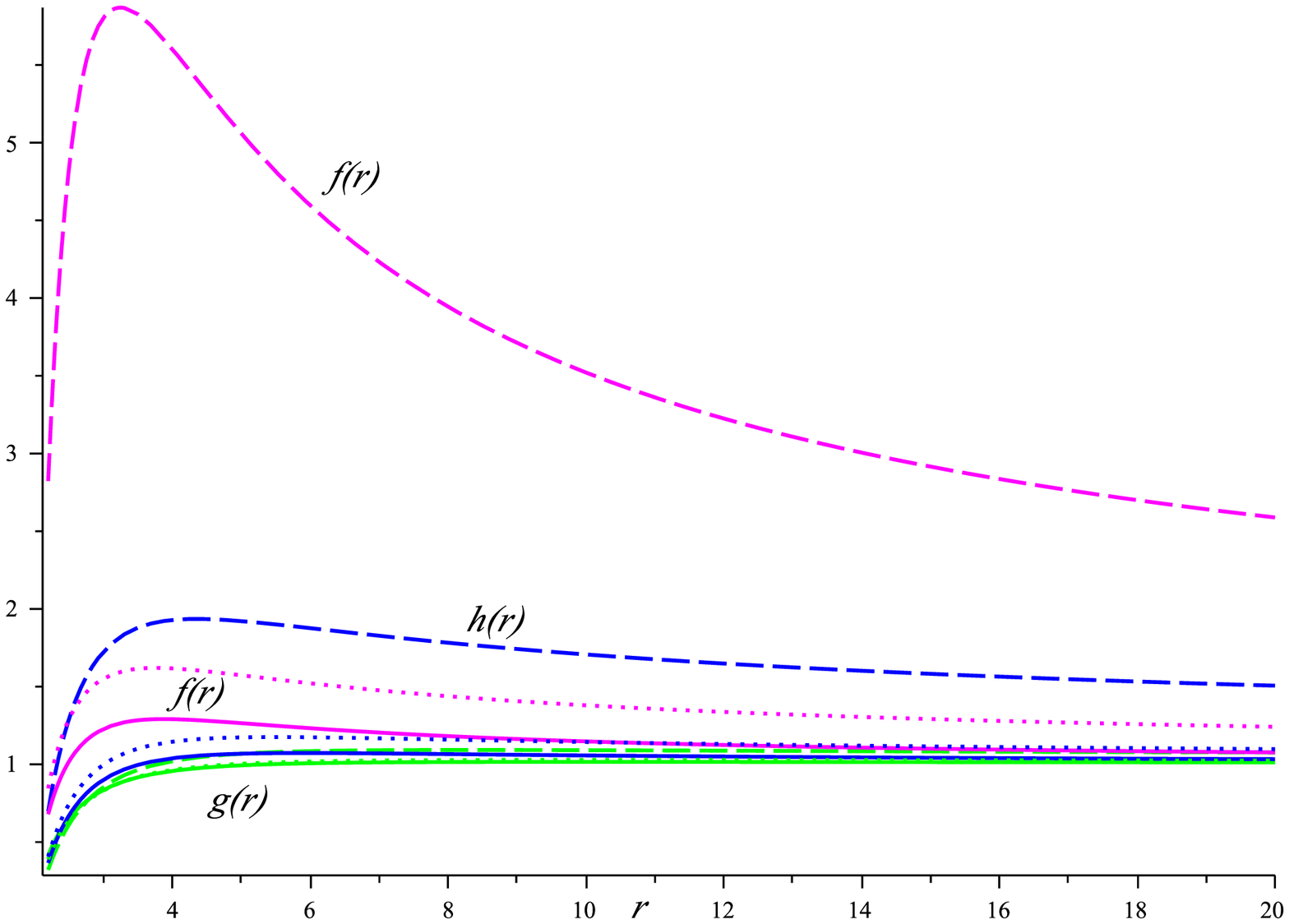}{Medium $z=2$ black hole ($r_0 = 2$) with $\mu=2.5$ and $\lambda=-10$ for $k=-1$ where Einsteinian gravity is solid, Gauss-Bonnet gravity is dashed, and quasi-topological gravity is dotted. Similar to above, $f(r0, g(r), h(r)$ are magenta, green, and blue.}{largemu}{0.4}

\section{Thermodynamics}
In this section we generalize from $5$ to $D$ dimensions to study the thermodynamic
behaviour of the solutions we obtain. The Iyer/Wald  prescription for black hole
entropy is \cite{Iyer}
\begin{equation}
S = -2 \pi \oint d^{D-2} x \sqrt{\tilde{g}} Y^{a b c d} \hat{\epsilon}_{ab} \hat{\epsilon}_{cd}, \hspace{5mm} \text{where} \hspace{2mm} %
Y^{a b c d} = \frac{\partial{\mathcal{L}}}{\partial{R_{a b c d}}}
\end{equation}
where $\hat{\epsilon}_{a b}$ is the binormal to the horizon and $\mathcal{L}$ is the Lagrangian, with Latin indices denoting quantities projected onto the horizon surface.
For the static black holes
considered here, $Y = Y^{a b c d} \hat{\epsilon}_{ab} \hat{\epsilon}_{cd}$ is constant on the horizon and so the entropy is given
simply as
\begin{equation}
S=-2\pi Y \int d^{D-2}x\sqrt{\tilde{g}},
\end{equation}
where the integration is done on the $(D-2)$-dimensional spacelike
hypersurface of the Killing horizon with induced metric $\tilde{g}_{a b} $ (whose determinant is $\tilde{g}$).
Although the asymptotic behaviour of our solution is different from Ref. \cite{Myers} and $f(r) \neq g(r)$, we obtain the same result:
\begin{equation}
S_k = \frac{A}{4 G_D} \left( 1 + \frac{2 (D -2)}{D - 4}\lambda k \frac{L^2}{r^2_0} -%
 \frac{3\left( D - 2 \right) }{D - 6} \mu k^2 \frac{L^4}{r^4_0} \right)
\label{entropy_general}
\end{equation}
where $D$ is the number of dimensions and $A$ is the surface area of the black hole (since our metric is spherically symmetric, the surface area will be proportional to $r^{D-2}_0$).

The temperature of the black holes is found by ensuring regularity at the horizon after Wick-rotation; we obtain
\begin{equation}
T = \left( \frac{r^{z+1} \sqrt{f^{\prime}g^{\prime}}}{4 \pi L^{z+1}}\right)_{r=r_0} .
\end{equation}
This quantity can be numerically calculated, and plotted against entropy on a logarithmic scale, to study stability of the black holes.
A negative slope indicates that the black hole will not be in thermal equilibrium and so must decay.

\subsection{Stability of AdS Black Holes}

Plotting the solution for $z=1$ in five dimensions, which can be checked with the analytic case,
we obtain Figure \ref{z1ThermoNum}, where we use $\lambda = 0.4$ and $\mu = -0.001$.
The solid line is quasi-topological gravity, while dots correspond to Gauss-Bonnet and crosses are Einsteinian.
The parameter $k$ varies between $-1, \, 0, \, 1$, coloured green, blue, and magenta, respectively.
Up to the black hole sizes for which we are able to find valid numerical solutions,  we see no evidence of unstable black holes for any value of $k$.
In the Einsteinian case
we see that small black holes will become unstable for $k=1$, so it is expected that for sufficiently small values of $\lambda$ and $\mu$, the solution will be one of small, unstable black holes.

To see what effect the sign of the quasi-topological parameter has on black hole stability, we plotted a similar set of curves for a positive value of $\mu=0.001$.  This plot is shown in Figure \ref{z1ThermoPos}.
We see that for $k=+1$ sufficiently small black holes are thermodynamically unstable.
\mapleplot{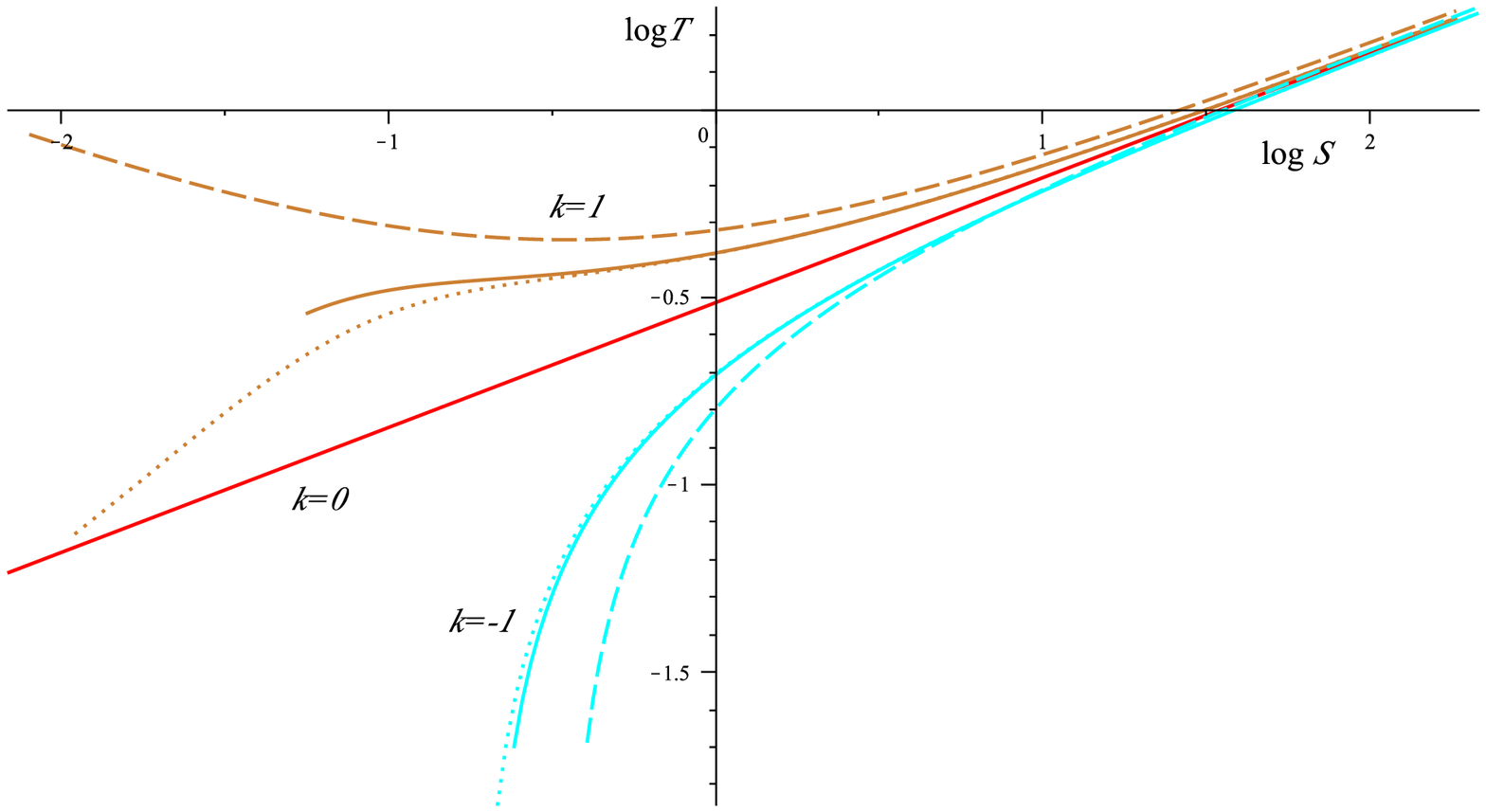}{log($T$) versus log($S$) for $z=1$ , $\lambda = .04$, and $\mu = -.001$.  The solid line is quasi-topological gravity, the dotted Gauss-Bonnet, the dashed Einsteinian. The parameter $k$ varies between $-1, \, 0, \, 1$, coloured turquoise, red, and brown, respectively.}{z1ThermoNum}{0.5}
\mapleplot{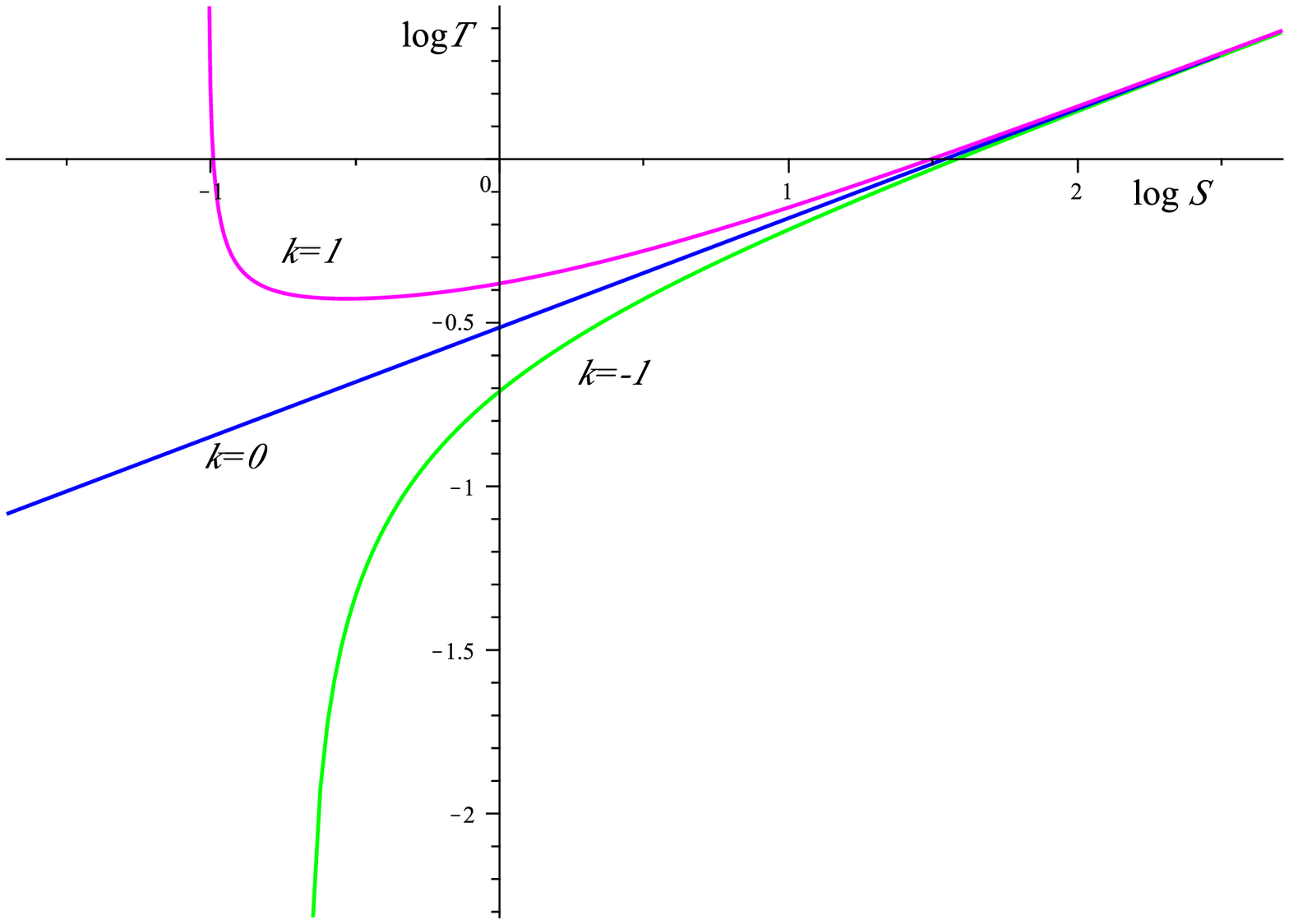}{log($T$) versus log($S$) for $z=1$, $\lambda=.04$, and $\mu=0.001$. The parameter $k$ varies between $-1, \, 0, \, 1$, coloured green, blue, and magenta, respectively.}{z1ThermoPos}{0.4}

\subsection{Stability of Lifshitz Black Holes}

For $z=2$, we also plot log($T$) versus log($S$).
Note that this plot is also specific to the five dimensional case.
For large black holes, it appears that by varying $z$ we do not change the temperature-entropy relationship, but instead merely introduce a scaling factor to both entropy and temperature terms.
It is also apparent that in both cases, positive Gauss-Bonnet and quasi-topological terms will both introduce $k=+1$ instability in black hole solutions.
A sufficiently negative quasi-topological term is also seen to partly counteract the positive Gauss-Bonnet term in Figure \ref{z2ThermoNum}.

\mapleplot{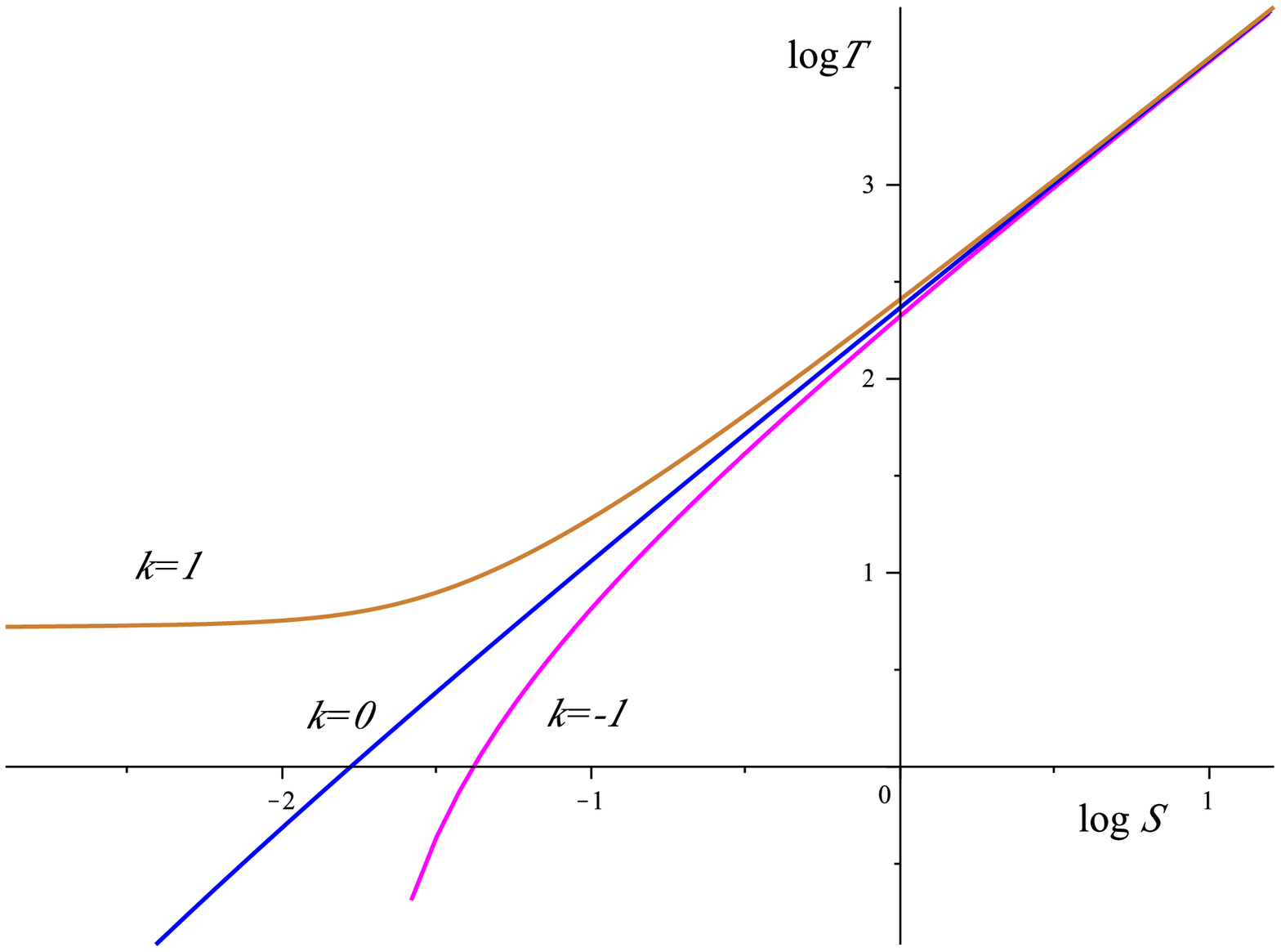}{log($T$) versus log($S$) for $z=2$, $\lambda=.04$, and $\mu=-0.001$. The parameter $k$ varies between $-1, \, 0, \, 1$, coloured magenta, blue, and brown, respectively. }{z2ThermoNum}{0.4}
\mapleplot{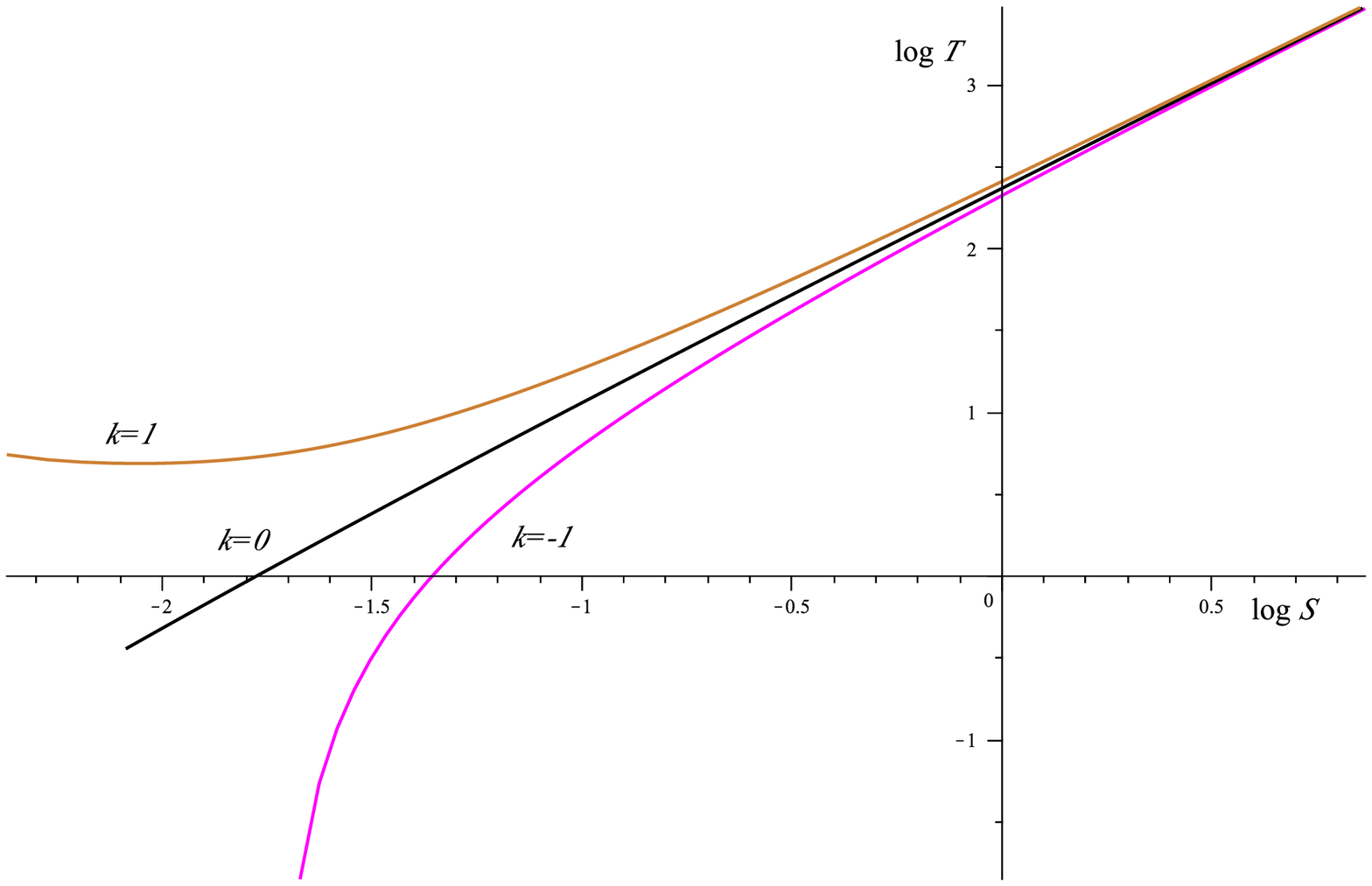}{log($T$) versus log($S$) for $z=2$, $\lambda=.04$, and $\mu=-.0003$. The parameter $k$ varies between $-1, \, 0, \, 1$, coloured magenta, black, and brown, respectively.}{z2ThermoCurved}{0.5}

\section{Conclusions}

 It is well-known that the third order Lovelock term (cubic in the Riemann tensor) does not appear in the field equations in five dimensions as it is a topological invariant.  Terms cubic in curvature in general yield higher-order differential equations for metric components.  Quasi-topological gravity \cite{Myers} is an exception to this general rule -- the cubic terms conspire to yield second-order differential equations for spherically symmetric metrics.

The main result of our paper is to demonstrate that a broad class of solutions -- those that are asymptotic to Lifshitz gravity -- exist in quasi-topological gravity in five dimensions.  We obtain a family of solutions dependent on two parameters, one giving a measure of the gauge field strength and the other the black hole radius.  For a given value of the gauge field strength, we found that there exists a unique solution with asymptotic Lifshitz behaviour. Varying the gauge field strength, we found that there exists a family of solutions for a given black hole radius. The $r$-dependence of these solutions varies considerably: the metric functions can develop a ``spike"  by increasing the gauge field strength.
We also find that in general, the quasi-topological term acts similarly to the Gauss-Bonnet term, but in negative sign.
When a negative Gauss-Bonnet term decreases the magnitude of the spike, a positive quasi-topological term will have the same effect.
We see this when a positive quasi-topological parameter is added to a Gauss-Bonnet solution that has decreased the magnitude of an Einsteinian spike in $g(r)$: our quasi-topological parameter further decreases the magnitude of the spike.

We also investigated the thermal stability of these quasi-topological Lifshitz black holes.
We found that a negative quasi-topological term, just like a positive Gauss-Bonnet term, will prevent instabilities in what are ordinarily unstable Einsteinian black holes.
For the asymptotically AdS case ($z=1$) we found that
for sufficiently negative values of $\mu$ the instabilities that arise in Einsteinian gravity may be removed, in the same way that a sufficiently positive Gauss-Bonnet term removes the small-$r$ black hole instability.
The AdS solutions were seen to be unstable for positive values of $\mu$.
With regard to the stability of asymptotically Lifshitz solutions with $z=2$ in Einstein gravity \cite{Mann}, we found that the quasi-local term with positive coupling constant
furnishes an effect similar to a positively coupled Gauss-Bonnet term.

It is clear that there is much to explore in Lifshitz gravity with the addition of higher-order curvature corrections.  The implications of
these corrections for the dual theory remain an interesting subject for future investigation.

\appendix
\section{The conserved quantity along the radial coordinate $r$ \label{Conserved}}
In this appendix, we demonstrate the existence of a constant $\mathcal{C}_{0}$, which
conserved along the radial coordinate $r$, and compute its form. Since there is no exact Quasi-topological-Lifshitz solution (except under special circumstances), we calculate it at the horizon and at infinity.

Reparametrizing the metric with the relations
\begin{align}
F(r) &= \frac{1}{2} \ln{f(r)} + z \ln{\frac{r}{L}}, \nonumber \\
G(r) &= -\frac{1}{2} \ln{g(r)} - \ln{\frac{r}{L}}, \nonumber \\
R(r) &= \ln{\frac{r}{L}}, \nonumber \\
H(r) &= \ln{h(r)} + z \ln{\frac{r}{L}},
\end{align}
the metric becomes
\begin{equation}
ds^2 = -e^{2 F(r)} dt^2 + e^{2 G(r)} dr^2 + e^{2 R(r)} \frac{1}{L^2} d\Omega^2
\end{equation}
whose form we insert into the action.
Following a similar method to reference \cite{Deh3}, we integrate by parts and obtain a one-dimensional Lagrangian $\mathcal{L}_{1D} = \mathcal{L}_{1g} + \mathcal{L}_{1m}$ as
\begin{align}
\mathcal{L}_{1g} &= (D-2) \left( -2 \frac{\Lambda}{D-2}e^{2G} + \left[ 2 F^{'} R^{'}%
+ (D-2) R^{\prime 2} \right] \right. - \frac{\lambda L^2}{3} \left[ 4F^{\prime} R^{\prime 3} +%
(D-5)R^{\prime 4}\right] e^{-2G} \nonumber \\
& \quad - \left.\vphantom{\frac{\Lambda}{D-2}}%
\frac{\mu}{5} L^4 \left[ 6F^{\prime}R^{\prime 5} + (D-7)R^{\prime 6} \right]%
e^{-4 G} \right) e^{F - G + (D-2) R} \nonumber \\
\mathcal{L}_{1m} &= \frac{1}{2}q^2 \left( m^2 + H^{\prime 2} e^{-2G} \right)%
e^{-F + G + (D-2)R + 2H}.
\end{align}

We are then able to write the equations of motion in the same manner as \cite{Deh3}, obtaining the conserved quantity
\begin{align}
\mathcal{C}_0 &= 2\left(F^{\prime}-R^{\prime}\right) \left(1 - 2\lambda L^2 R^{\prime%
2}e^{-2G} - 3 \mu L^4 R^{\prime 4}e^{-4G} \right) e^{F - G + (D-2)R} \nonumber \\
& \quad - q^2 H^{\prime}e^{-F -G +(D-2)R + 2H} \nonumber \\
&= \left[ \left( 1 - 2 \lambda g - 3 \mu g^2 \right) \left( r f^{\prime} + 2 \left( z - 1 \right) f \right) - q^2 \left( z h + r h^{\prime} \right) h%
\right] \frac{r^{z+D-2}}{L^{z+1}} \left( \frac{f}{g} \right)^{1/2}
 \label{Constant}
\end{align}
This derivation was performed using the form of the quasi-topological Lagrangian for $D$ dimensions, given by (\ref{quasitop}), and the form of the conserved quantity was checked explicitly for $D$ from $7$ through $11$ to determine the dimensionally independent form given. For any value of $z$, this conserved quantity arises from the symmetry
\begin{equation}
\left(
\begin{array}{c}
F(r) \\
R(r) \\
G(r) \\
H(r)
\end{array}
\right) \to \left(
\begin{array}{c}
F(r) + \delta \\
R(r) - \frac{\delta}{D-2} \\
G(r) \\
H(r) + \delta
\end{array} \right)
\end{equation}
For $z=1$, $f(r) = g(r)$ and the constant  reduces to
\begin{equation*}
\mathcal{C}_0 = \frac{r^{D}}{L^2} \left( f - \lambda f^2 - \mu f^3 \right)^{\prime}
\end{equation*}

\section{Near-horizon Series Solution Coefficients \label{coeffs}}

Here we write down the remaining coefficients of the near-horizon series solution (\ref{near-hor})  up to second order. Defining for simplicity $L_0 = -1 + 2\lambda + 3\mu$, we
obtain
\begin{align*}
f_2 &= \left( -6 g_1 {r_0}^8 \left( z \left( \lambda g_1 - \frac{2}{3} {h_1}^2 L_0 \right) + \frac{2}{3} {h_1}^2 L_0 \right) +%
	12 z {r_0}^7 \left( z \left( g_1 + \frac{1}{2} L_0 {h_1}^2 \right) + \frac{3}{4} g_1 - \frac{1}{4} L_0 {h_1}^2 \right) \right. \\
    & \quad \left. + 18 z {r_0}^6 \left( \frac{1}{9} L_0 z^2 + \frac{2}{9} L_0 z + \frac{1}{3} L_0 + \mu L^2 k {g_1}^2 \right) +%
	24 k L^2 g_1 \lambda z {r_0}^5 \left( z - \frac{1}{4} \right) - 36 k^2 L^4 g_1 z \mu {r_0}^3 \left( z - \frac{5}{4} \right) + 24 z \mu L^6 k^3 \right) \\
    & \hspace{7mm} \left( r_0 \left( L_0 {h_1}^2 g_1 {r_0}^8 - 9 z {r_0}^7 g_1 - z {r_0}^6 \left( L_0 z^2 + 2 L_0 z + 3 \mu + 6 \lambda - 9 \right) \right. \right. \\
    & \hspace{8mm} \left. \left. - 18 z \lambda L^2 k g_1 {r_0}^5 + 6 z L^2 k {r_0}^4 + 27 L^4 k^2 \mu z {r_0}^3 g_1 + 6 z \mu L^6 k^3 \right) \right)^{-1} \\
g_2 &= \left( 2 g_1 {r_0}^{12} \left( L_0 \left( g_1 + {h_1}^2 \left( z^2 + 1 \right) L_0 \right)  \right. \right. \\
    &  \left. \quad + z \left( -3 {g_1}^2 \lambda - 2 {h_1}^2 g_1 L_0 - 2 {h_1}^4 {L_0}^2 \right)%
	\right)  -3 z g_1 {r_0}^{11} \left( z \left( 2 g_1 + 5 L_0 {h_1}^2 \right) - 13 g_1 + 5 {h_1}^2 L_0 \right) +  \\
    & \quad + {r_0}^{10} \left( -2 z^4 L_0 \left( g_1 + {h_1}^2 L_0 \right) + 2 z^3 L_0 \left( 2 g_1 - {h_1}^2 L_0 \right) \right. \\
    & \quad + \left. z^2 \left( 4 L_0 L^2 k \lambda {h_1}^2 {g_1}^2 + %
    	g_1 \left( 42 \mu + 20 \lambda + 2 \right) + 2 {h_1}^2 L_0 \left( 3 \mu - 2 \lambda + 7 \right) \right) \right. \\
    & \quad \left. + z \left( 6 {g_1}^3 L^2 k \left( 3 \mu - 2 \lambda^2 \right) - 8 L^2 k \lambda {h_1}^2 {g_1}^2 L_0 + g_1 \left(24 \mu + 48 \lambda - 72 \right) +%
    	6 {h_1}^2 L_0 \left( \mu + 2 \lambda - 3 \right) \right) + 4 L^2 k \lambda {h_1}^2 {g_1}^2 L_0 \right) \\
    & \quad - 6 L^2 k z g_1 \lambda {r_0}^9 \left( z \left( 4 g_1 + 5 {h_1}^2 L_0 \right) - 22 g_1 + 5 {h_1}^2 \right) \\
    & \quad - 4 L^2 k {r_0}^8 \left( z^4 \lambda g_1 L_0 - 2 z^3 \lambda g_1 L_0 + z^2 \left( \frac{3}{2} \mu L^2 k {h_1}^2 {g_1}^2 L_0 + g_1 \left( -3 - \lambda - 10%
    	\lambda^2 - 21 \lambda \mu \right) - 3 {h_1}^2 L_0 \right) \right. \\
    & \left. \quad + z \left( - \frac{27}{2} k L^2 {g_1}^3 \lambda \mu - 3 L_0 \mu L^2 k {h_1}^2 {g_1}^2 + 3 g_1 \left( -8 \lambda^2 - 4 \mu \lambda + 12 \lambda + 3 \right) %
    	+ 3 {h_1}^2 L_0 \right) + \frac{3}{2} L_0 \mu L^2 k {h_1}^2 {g_1}^2 \right) \\
    & \quad + 9 L^4 z k^2 g_1 {r_0}^7 \left( z \left( g_1 \left( 4 \mu - \frac{8}{3} \lambda^2 \right) + 5 \mu {h_1}^2 L_0 \right) + g_1 \left( -18 \mu + 12 \lambda^2 \right)%
    	- 5 \mu {h_1}^2 L_0 \right) \\
    & \quad + 6 L^4 z k^2 g_1 {r_0}^6 \left( z^3 \mu L_0 - 2 z^2 \mu L_0 + z \left( -21 \mu^2 - \mu - 10 \lambda \mu + 4 \lambda \right) - 9 k L^2 \mu^2 {g_1}^2 - 12 \mu^2%
    	+ \mu \left( 36 - 24 \lambda \right) - 12 \lambda \right) \\
    & \quad + 36 \mu L^6 z k^3 {g_1}^2 \lambda {r_0}^5 \left( 2 z - 7 \right) - 12 \mu L^6 z k^3 {r_0}^4 \left( z \left( 2 g_1 - L_0 {h_1}^2 \right)%
    	- 8 g_1 + {h_1}^2 L_0 \right) \\
    & \quad \left. + 27 \mu^2 L^8 z k^4 {g_1}^2 {r_0}^3 \left( 2 z - 5 \right) + 24 z g_1 \mu L^8 k^4 \lambda {r_0}^2 \left( z - 1 \right) - %
    	36 z g_1 \mu^2 L^10 k^5 \left( z - 1 \right) \right) \\
    & \cdot \left( {r_0} \left( - {r_0}^4 - 2 \lambda L^2 k {r_0}^2 + 3 L^4 k^2 \mu \right) %
    	\left( - L_0 g_1 \left( z - 1 \right) {h_1}^2 {r_0}^8 + 9 z g_1 {r_0}^7 \right. \right. \\
    & \quad + z {r_0}^6 \left. \left. \left( z^2 L_0 + 2 z L_0 + 3 \mu - 9 + 6 \lambda \right) + 18 z \lambda L^2 k g_1 {r_0}^5 - 6 z L^2 k {r_0}^4 - %
    	27 L^4 k^2 \mu z {r_0}^3 g_1 - 6 z \mu L^6 k^3 \right) \right)^{-1} \\
h_2 &= - h_1 \left( {r_0}^5 \left( z \left( - 2 g_1 - {h_1}^2 L_0 \right) - 3 g_1 + {h_1}^2 L_0 \right) + 3 z {r_0}^4 - 4 L^2 k g_1 %
	\lambda {r_0}^3 \left( 2 z + 3 \right) \right. \\
    & \left. \quad + 6 z \lambda L^2 k {r_0}^2 + 3 \mu L^4 k^2 g_1 r_0 \left( 2 z + 3 \right) - 9 L^4 k^2 \mu z \right) \\
    & \quad \cdot \left( {r_0}^2 g_1 \left( -2 {r_0}^4 - 4 \lambda L^2 k {r_0}^2 + 6 L^4 k^2 \mu \right) \right)^{-1}
\end{align*}
where $g_1$ is given by (\ref{g1-nearhor}).  Each coefficient depends on the independent parameters
$r_0$ (the horizon radius) and $h_1$ (proportional to the field strength at the horizon).

\section{Large $r$ Series Solutions \label{larger}}

For large distances away from the black hole, $r>>L$, we present here series solutions for $5$ dimensions.

Considering first the ansatz in (\ref{larger_ansatz})  for $k=0$, the
field equations to first order of $\varepsilon$ imply
\begin{align*}
0 &=  2 r^2 {h_e}^{\prime \prime} + 2 r {h_e}^{\prime} \left( z + 4 \right) + z r \left( {g_e}^{\prime} - {f_e}^{\prime} \right) + 6 z g_e \\
0 &=  2 r \left( z - 1 \right) {h_e}^{\prime} + 3 r {g_e}^{\prime} + \left( z^2 - z + 12 \right) g_e + \left(z + 3 \right) \left( z - 1 \right) \left( 2 h_e - f_e \right) \\
0 &=  2 r \left( z - 1 \right) {h_e}^{\prime} + 3 r {f_e}^{\prime} + \left( z^2 + \frac{51\mu + 22 \lambda - 5}{3\mu + 2\lambda - 1} z - 18\mu - 6 \right) g_e + \left( z - 3 \right) \left( z - 1 \right) \left( 2 h_e - f_e \right)
\end{align*}
The solutions for $f_e(r), g_e(r), \mbox{ and } h_e(r)$ yield integer powers of $r$ in a number of special cases.  For $h_e(r)$, we find
\begin{align}
h_e(r) &= C_1 r^{-3 - z} + r^{\left( -3 - z\right) / 2} \left( C_2 r^{-\gamma/2} + C_3 r^{\gamma/2} \right) \\
f_e(r) &= \mathcal{D}_1 r^{-3 - z} + \mathcal{K} r^{\left( -3 - z\right) / 2} \left( \mathcal{D}_2 r^{-\gamma/2} + \mathcal{D}_3 r^{\gamma/2} \right) \\
g_e(r) &= \mathcal{D}_1 r^{-3 - z} + \mathcal{K} r^{\left( -3 - z\right) / 2} \left( \mathcal{F}_2 r^{-\gamma/2} + \mathcal{F}_3 r^{-\gamma/2} \right)
\end{align}
where $C_1$, $C_2$, and $C_3$ are integration constants, $L_0$ is defined as in Appendix \ref{coeffs}, and
\begin{align*}
\gamma^2 &= \left\{ \left( -21 \mu + 2 \lambda - 9 \right) z^2 + \left( 18 \mu - 20 \lambda + 26 \right) z + 51 \mu + 50 \lambda - 33 \right\} \left( -1 + 3 \mu + 2 \lambda \right)^{-1} \\
\mathcal{D}_1 &= C_1 \frac{ 2 L_0 \left( z - 3 \right) \left( z-1 \right)}{ 2 L_0 z^2 - \left( 1 + 6 \lambda + 21 \mu \right) z - \left( 9 + 6 \lambda + 45 \mu \right)} \\
\mathcal{D}_2 &= C_2 \left( \mathcal{L}_1 + \mathcal{L}_2 \right) \\
\mathcal{D}_3 &= C_3 \left( \mathcal{L}_1 - \mathcal{L}_2 \right) \\
\mathcal{F}_2 &= C_2 \left( \mathcal{L}_3 + \mathcal{L}_4 \right) \\
\mathcal{F}_3 &= C_3 \left( \mathcal{L}_3 - \mathcal{L}_4 \right) \\
\mathcal{L}_1 &= \left( \left(1-8\lambda-21\mu \right)z + 2 + 2\lambda + 12\mu \right)%
		\left( \cdot \left(36 \lambda^2 + \left( 132\mu - 28\right) \lambda +9 - 30\mu +153\mu^2\right)z^2 \right. \\
	& \quad	\left. +\left(-40\lambda^2 + \left(-72\mu + 56\right) \lambda + 12\mu - 90\mu^2 - 26\right) z + 68\lambda^2 + \left(-92 + 132\mu \right) \lambda %
		-78\mu + 33 +81\mu^2 \right)^{1/2} \\
\mathcal{L}_2 &= \left( -8\lambda^2 + \left(-10 -78\mu \right) \lambda -153\mu^2 -24\mu + 1\right) z^2 + \\
		& \quad \left( 20\lambda^2 + \left( 132\mu + 4\right) \lambda + 5 + 261\mu^2 + 42\mu \right) z + 6\left( 6\lambda + 16\mu - 1\right) \left( \lambda - 1 \right) \\
\mathcal{L}_3 &= \left( z - 1 \right) \cdot L_0 \cdot 
		\left( \cdot \left(36 \lambda^2 + \left( 132\mu - 28\right) \lambda +9 - 30\mu +153\mu^2\right)z^2 \right. \\
	& \quad	+\left(-40\lambda^2 + \left(-72\mu + 56\right) \lambda + 12\mu - 90\mu^2 - 26\right) z + 68\lambda^2 + \left(-92 + 132\mu \right) \lambda %
		\left. -78\mu + 33 +81\mu^2 \right)^{1/2} \\
\mathcal{L}_4 &= \left( z - 1 \right) \cdot L_0 \cdot \left( \left( -1 - 18 \lambda - 21\mu \right) - 9 \mu - 10 \lambda + 27 \right) \\
\mathcal{K} &= \frac{1}{2 z \left( 12 z \mu - z + 5 z \lambda + \lambda -2 - 3 \mu \right) L_0} \\
\end{align*}

For $k = \pm 1$, we can represent the asymptote functions as series
\begin{align*}
f(r) &= 1 + \sum_{i=1}^{2 (n  + z) - 3} \frac{a_i}{r^i} \\
g(r) &= 1 + \sum_{i=1}^{2 (n  + z) - 3} \frac{b_i}{r^i} \\
h(r) &= 1 + \sum_{i=1}^{2 (n  + z) - 3} \frac{c_i}{r^i} .
\end{align*}
The coefficients can be determined from direct calculation. Due to the equivalence of our field equations, they match the values obtained in third order Lovelock gravity \cite{Deh2}, once the substitutions $\mu L^4 =-\hat{\alpha}_3$ and $\lambda L^2 =\hat{\alpha}_2$ are made.

For $z=2$, we obtain nonzero coefficients only for the powers $r^{-2}, r^{-5}, r^{-7}, r^{-9}$.
Just as in Lovelock gravity, at $z=2$, all of the even powers of $r$ until $r^{-5}$ are present in the large $r$ expansion of the asymptote functions.

\acknowledgements
This work was supported by the Natural Sciences and Engineering Research Council of Canada and Research Institute for Astrophysics and Astronomy of
Maragha.

\end{document}